\documentclass[epj,final]{svjour}

\usepackage{latexsym}
\usepackage{url}
\usepackage{amsfonts}
\usepackage{amssymb}
\usepackage{amsmath}
\RequirePackage{graphicx}

\begin{document}
\title{Cosmological constant as a fundamental constant}
\author{V.G. Gurzadyan\inst{1,2},
A. Stepanian\inst{1}
}                     
%
%
\institute{Center for Cosmology and Astrophysics, Alikhanian National Laboratory and Yerevan State University, Yerevan, Armenia \and SIA, Sapienza Universita di Roma, Rome, Italy
}
\date{Received: date / Revised version: date}
%

\abstract{We consider further consequences of recently \cite{GS1} revealed role of cosmological constant $\Lambda$ as of a physical constant, along with the gravitational one to define the gravity i.e. the General Relativity and its low-energy limit.  We now show how $\Lambda$-constant affects the basic relations involving the Planck units and leads to emergence of a new dimensionless quantity (constant) which can be given cosmological information content. Within Conformal Cyclic Cosmology this approach implies the possibility of rescaling of physical constants from one aeon to another; the rescaling has to satisfy a condition involving $\Lambda$ and admitting group symmetry. The emerged dimensionless information constant enables to reduce the dynamics of the universe to an algorithm of discrete steps of information increase.} 
\PACS{
      {98.80.-k}{Cosmology}   
     } 

%
\maketitle
\section{Introduction}

The cosmological constant as a universal constant was mentioned by Einstein \cite{E,E1} when he introduced it to describe a static cosmological model but later abandoned it. In the recent study \cite{GS1} we have shown that the cosmological constant $\Lambda$ does possess properties of a second physical constant, along with the gravitational one $G$, both for the General Relativity and the Newtonian gravity as its low-energy limit. That follows from
the Newton theorem on the equivalency of gravity of a sphere and of a point-mass located in its center (see \cite{G85}), i.e. the motivation of emergence of $\Lambda$ is entirely different from that of the static universe. The $\Lambda$-constant
was shown to be dimension-independent and matter uncoupled and therefore even more universal than the gravitational
constant \cite{GS1}. That approach enables one to describe the dark matter and the dark energy as possessing common nature \cite{G}.  

Here we discuss further consequences for $\Lambda$ joining the set of fundamental constants $G, c, \hbar$, the gravitational (Newton) constant, speed of light and Planck constant, respectively; for detailed discussion of constants see \cite{U}. 

The consideration of $\Lambda$ together with the 3 constants affects the issue of Planck units. Planck  \cite{PL} denoted the latter as natural units since they "retain their meaning for all times and for all cultures, even extraterrestrial and non-human ones".  
We show that $\Lambda$ together with the Planck units leads to emergence of a dimensionless constant, also relevant for "all cultures", which in cosmological context acts as scaling for information.  Among the consequences is that, a rescaling of values of the 3 fundamental physical constants will be allowed from one aeon to another aeon within the Conformal Cyclic Cosmology (CCC) \cite{P,GP}; the rescaling satisfies a dimensionless relation and possess group symmetry.    

\section{The 4 units and information}

The set of (3+1) constants and their units looks as
\begin{equation}\label{NatDim}
[c]=LT^{-1},\quad [G]=M^{-1}L^3T^{-2},\quad [\hbar]=ML^{2}T^{-1},\quad [\Lambda]=L^{-2},
\end{equation}
where  $L$, $T$ and $M$ stand for dimensionality of length, time and mass, respectively. The most general combinations of these constants can be represented in the form 
\begin{equation}\label{NatComb}
[c^{n_1}\Lambda^{n_2}G^{n_3}\hbar^{n_4}]=L^{n_1 -2n_2 + 3n_3 + 2n_4}T^{-n_1 - 2n_3 -n_4}M^{-n_3 + n_4}.
\end{equation}

From here, two consequences follow readily. 

First, for the set ($G, \Lambda, c, \hbar$) the corresponding algebraic equation has no unique solution and hence no units can be defined by these 4 constants, as distinct of the case of ($G, c, \hbar$) leading to Planck units.   

Second, the following dimensionless quantity (constant) does emerge
\begin{equation}\label{dimless}
I=\frac{c^{3a}}{\Lambda^a G^a \hbar^a},
\end{equation}
where $a$ is a real number. In contrast, no dimensionless quantity was possible to construct from the 3-set ($G, c, \hbar$). This difference, as we show below, can have consequences for the CCC.

Note, that for $a=1$ in (\ref{dimless}) one has $I \simeq 3.4 \times  10^{121}$, which obviously reflects the renown cosmological constant value problem.   

The relation of the 4 constants in (\ref{dimless}), except for a numerical factor, for $a=1$ coincides with that of the information (or entropy, with the Boltzmann constant) of de Sitter event horizon \cite{Bek,HG,P}
\begin{equation}\label{dSE}
I_{dS}= 3 \pi \frac {c^3}{\Lambda G \hbar}.
\end{equation}
This relation emerges also from the Bekenstein Bound \cite{BekB} written for the information in de Sitter space 
\begin{equation}\label{BBdS}
I_{BB} =  \frac {3 \pi c^3}{\Lambda G \hbar ln 2}.
\end{equation}
One may expect emergence in future of this same dimensionless relation of the 4 constants from other backgrounds or motivations.  
 
The coincidence of $I_{DS}$ and $I_{BB}$ i.e. $\Delta I_{dS}=0$, reflects that there is no  information (entropy, thermodynamical) time evolution in de Sitter manifold. In the next section we will study the possible link between this statment and symmetries of manifolds with more details.

The importance of Newton theorem lies also on the fact that it enables one to generalize the ``sphere-point" equivalence idea to higher dimensions. In those cases, of course we have hyperspheres $\mathbb{S}^{d-1}$, where $d$ is the dimensionality of space. Then, according to \cite{GS1}, for gravitational potential (d$\geq$3) we have
\begin{equation}\label{PotN}
\Phi(r)=-\frac{G_{d} M}{r^{d-2}}-\frac{\Lambda c^2 r^2}{2d}.
\end{equation} 
As a consequence, the Newton gravitational constant becomes dimension-dependent and for four constants we have
\begin{equation}\label{NatDim}
[c]=LT^{-1},\quad [G_{d}]=M^{-1}L^dT^{-2},\quad [\hbar]=ML^{2}T^{-1},\quad [\Lambda]=L^{-2},
\end{equation}
Then, the dimensionless quantity is obtained as

\begin{equation}\label{dimlessd}
I_{d}=\frac{c^{3a}}{G^a_{d} \hbar^a \Lambda^{a \frac{d-1}{2}} }, \quad a \in \mathbb{R},
\end{equation}
obviously, for $d=3$ we recover Eq.(\ref{dimless}).

Since the information is related to the area of (d-1)-dimensional hypersurface, the Bekenstein's ``elementary particle" has an area
\begin{equation}
\frac{d\pi^{\frac{d}{2}-1}}{\Gamma(\frac{d}{2}+1)} \frac{G_d \hbar}{c^3},
\end{equation} 
and the evolution of the universe ends at de Sitter phase at corresponding maximum information
\begin{equation}
I_{dS} =\frac{c^3}{G_d \hbar  \Lambda^{\frac{d-1}{2}}} d^{\frac{d-1}{2}} \pi \,\, bits.
\end{equation}

\section{Information, time evolution and Weyl principle}

As shown above, upon introducing $\Lambda$ as one of fundamental constants the notions of $l_p$, $m_p$ and $t_p$ as ordinary natural units, disappear. However, within Bekenstein's ``elementary particle" \cite{Bek} approach, one can consider Planck units as composing one bit of information. Namely, one bit of information is attributed to $4 l_p^2$, so that in expanding universe upon the increase of the surface area more information is created. Creation of information continues until in de Sitter (dS) phase $I_{dS}$-th bit is created.

Thus the time evolution of the universe is reduced to discretized steps
\begin{equation}
   T= \left\{1, 2, 3,....,I_{dS}\right\}.
\end{equation}

Such description based on creation of information, naturally, imposes a temporal order. Note, that for de Sitter (dS) universe where $\Delta I_{dS}=0$, we have time-translational symmetry T(t) as the subgroup of isometry group O(1,4). Thus it seems that, there might be a link between the T(t) group and evolution of universe based on information.

\begin{table}
\caption{}\label{tab4}
\centering
\begin{tabular}{ |p{1cm}||p{2.1cm}|p{4.7cm}|p{1cm}|p{2.5cm}|}
\hline
\multicolumn{5}{|c|}{Non-Relativistic Background Geometries} \\
\hline
Sign& Geometry&Symmetry Group&T(t)& Relativistic limit\\
\hline
$\Lambda > 0$ &Newton-Hooke &NH${}^{+}$ (4)=(O(3) $\times$ O(1,1)) $\ltimes$ R${}^6$& O(1,1) & de Sitter \\
$\Lambda = 0$ & Galileo & Gal(4)=(O(3) $\times$ R) $\ltimes$ R${}^6$& R &Minkowski\\
$\Lambda <0 $ &Newton-Hooke &NH${}^{-}$(4)=(O(3) $\times$ O(2) ) $\ltimes$ R${}^6$& O(2) & Anti de Sitter\\
\hline
\end{tabular}
\end{table}

In non-relativistic limit the geometry of universe is considered as Galilean spacetime, which has 10-parameter symmetry group Gal(4), where time translations T(t) make a subgroup of Gal(4). This is also true at non-zero cosmological constant case, where the symmetry groups are NH${}^{\pm}$(4) as shown in Table 1. Meantime, for each case it is easy to show that there are non-relativistic limits of following groups
\begin{equation}
    O(1,4) \to (O(3) \times O(1,1)) \ltimes R^6,\quad O(2,3) \to (O(3) \times O(2) ) \ltimes R^6,\quad IO(1,3) \to (O(3) \times R) \ltimes R^6,
\end{equation}
where clearly there is again time-translational symmetry. This implies that it is not possible to fix a preferred direction of time based only on symmetrical features of background geometries for both relativistic and non-relativistic ones.

At the same time, following \cite{GS1}, the sphere-point identity implies that, at each point of background geometry (spatial), we have O(3) symmetry. As in all of these non-relativistic geometries the spatial algebra is Euclidean E(3)=O(3)$\ltimes$ R${}^{3}$, the Newton theorem is valid.

Thus, by considering Newton theorem and information theoretic evolution of universe simultaneously, it becomes clear that although the geometry initially during the creation of information does not posses T(t) symmetry group, O(3) is the stabilizer of the spatial geometry. All possible 3-geometries with O(3) as the stabilizer are listed in Table 2.

\begin{table}
\caption{}\label{tab4}
\centering
\begin{tabular}{ |p{2.2cm}||p{2.6cm}|p{1.4cm}|}
\hline
\multicolumn{3}{|c|}{Spatial Geometries} \\
\hline
Space& Symmetry Group &Curvature\\
\hline
Spherical: $\mathbb{S}^3$ &O(4) &+ \\
Euclidean: $\mathbb{R}^3$ & E(3) & 0 \\
Hyperbolic: $\mathbb{H}^3$ &O${}^{+}(1,3)$  &-\\
\hline
\end{tabular}
\end{table}

In relativistic cosmology the background geometry i.e. the Friedmann-Lemaître-Robertson-Walker (FLRW) metric is fixed by ``Weyl principle'' which assumes that at any moment of time the universe is homogeneous and isotropic
\begin{equation}\label{FLRW}
ds^2= c^2 dt^2 - a(t)^2 d\Sigma^2,
\end{equation} 
where $\Sigma$ is one of geometries listed in Table 2. Within the approach presented above, ``Weyl principle" becomes not just a matter of simplification, but a condition to have gravity satisfying Newton theorem, on one side, and enabling information theoretical consideration, on the other side.

\section{CCC: rescaling of physical constants}

The key elements of CCC are the Second law of thermodynamics and the positive $\Lambda$ \cite{P}. That naturally implies the involvement of physical constants and Planck units through the concepts of entropy and information. Namely, within CCC the initial point of each aeon corresponds to vanishing of Weyl tensor, ${\bf C}=0$, and then the evolution of each aeon is completed by de Sitter expansion.  The re-set of entropy at the conformal boundary of aeons is reached by the loss of information in massive black holes situated in galactic centers and Hawking evaporation.     

Since the expressions defined by 4 constants (\ref{dimless}) are dimensionless numbers, they are transformed identically from one aeon to another (regarding the information transfer to the next aeon see \cite{GP1}), as invariants with respect to conformal transformation
\begin{equation}
\tilde{g}_{\mu\nu}=\Omega^2 g_{\mu\nu}.
\end{equation}

Namely, the ratio 
\begin{equation}\label{quant}
    \frac{Q_{dS}}{Q_p}=m (\frac{c^3}{\hbar G \Lambda})^n = m I^n, \quad m,n \in \mathbb{R}
\end{equation}
of all physical quantities $\left\{Q \right\}$ in final (de Sitter) and initial (Planck) eras of an aeon will remain invariant under conformal transformations.

However, the invariance of $mI^n$ does not imply the invariance of each of 4 constants involved. In other words, the constants can be rescaled from one aeon to another
\begin{equation}
    c\to a_1 c, \quad \hbar \to a_2 \hbar, \quad G\to a_3G, \quad \Lambda \to a_4 \Lambda, \quad a_i \in \mathbb{R^{+}},
\end{equation}
keeping satisfied the condition
\begin{equation}
    \frac{a_1 ^3}{a_2 a_3 a_4}=1,
\end{equation}
From here we arrive at the conclusion that, the constants' transformations in an aeon are invariant under the following group

\begin{equation}
 S= \left \{ {\begin{pmatrix}
 a_{11}\quad 0 \quad  0 \quad  0  \\  
 0 \quad  a_{22} \quad0 \quad  0 \\ 
 0 \quad  0 \quad  a_{33} \quad  0\\  
 0 \quad  0 \quad  0 \quad  a_{44}
\end{pmatrix}},  \quad det|S|=1, \quad a_{11}=a_1^3, \quad a_{22}=a_2^{-1}, \quad a_{33}=a_3^{-1}, \quad a_{44}={a_{4}^{-1}} \right \}.
\end{equation}
This means that, the subsequent aeons can possess rescaling of constants $c, \hbar, G, \Lambda $ and of $Q_i$ keeping invariant the dynamics of an aeon.

Thus, the yet unknown ``Master Equation" of the universe has to admit the $S$ group's symmetry. 

Then, in view of the relation (\ref{quant}) we see that there is a noted difference between the role of $\Lambda$ and of other constants. In fact, since the $\Lambda$ is absent in Planck era scales, by fixing $\Lambda$'s value, the values of physical quantities (up to possible combination of other constants) $Q_{dS}$ are fixed. This property is possessed only by $\Lambda$, since fixing any of the rest three constants does not define either the initial or final stages of an aeon. If so, one can rewrite the group $S$ as follows

\begin{equation}
 S= \left \{ {\begin{pmatrix}
 q_{11}\quad0 \\ 
 0 \quad  q_{22}
\end{pmatrix}}, \quad q_{ii}\in \mathbb{R^+}, \quad det|S|=1 \right \},
\end{equation}
where in this case, $q_{11}=a_{11} a_{22} a_{33}$ and $q_{22}=a_{44}$.

Thus, the expansion of an aeon starts at positive $\Lambda$ and upon fixing its value, the values of 3 physical constants $c, G$ and $\hbar$ are fixed according to formula (\ref{dimless}), i.e. allowing several equivalent combinations satisfying the group $S$.

Note, a difference of the described information approach and the conventional one defining the dynamics of the universe with Friedmannian equations.  Those equations are solved numerically for given input parameters with proper choice of time steps.  
Now, when $\Lambda$ is considered a universal constant and the notion of natural units for time, length and mass disappear,  we come to a dimensionless information and the dynamics of the universe is reduced to discrete steps $\left\{1, 2, 3,....,I_{dS}\right\}$.

\section{Conclusions}

The cosmological constant $\Lambda$, which as shown in \cite{GS1}, acts as a physical constant defining the gravity, in combination with other fundamental constants leads to the following principal conclusions:

a) the 4 constants no longer define a unique scaling for length, time and mass, as were the Planck units for the 3 physical constants;  

b) a dimensionless quantity (constant) is emerging composed of 4 constants $G, \Lambda, c, \hbar$ of a transformation group symmetry which was not possible with 3 constants $G, c, \hbar$.

Starting from 1970s the notion of information as of dimensionless quantity was attributed to event horizons \cite{Bek,HG}. Now, as shown above, only together with $\Lambda$ one can construct a natural dimensionless quantity, to which within  Bekenstein's ``elementary particle"  approach one can attribute information content. 

Thus, $\Lambda$ as universal constant approach, enables one to consider dynamics of the universe as of (d+1)-dimensional Lorentzian geometry satisfying the following conditions:

$\bullet$ Newton theorem ensures O(d) symmetry at each point of d-dimensional spatial geometry;

$\bullet$ The evolution intrinsically imposes ``time ordering" as described by group theoretical analysis; 

$\bullet$ Evolution can be reduced to discrete increase of (dimensionless) information;

$\bullet$ Bekenstein's ``elementary particle'' corresponds to an area $\frac{d\pi^{\frac{d}{2}-1}}{\Gamma(\frac{d}{2}+1)} \frac{G_d \hbar}{c^3}$;

$\bullet$ Evolution tends to de Sitter phase with information $\frac{c^{3}}{G_{d} \hbar \Lambda^{\frac{d-1}{2}} } d ^{\frac{d-1}{2}}\pi $ bits.

The group properties of transformations involving the physical constants within the Conformal Cyclic Cosmology imply that, at any positive value of $\Lambda$ at initial state of each aeon the initial values of 3 physical constants will allow rescaling satisfying the dimensionless constraint. The rescaling of 4 fundamental constants will admit the same global cosmological dynamics but with rescaled internal physics. This opens an entire arena for modifications for physical processes and configurations from one aeon to another, since the values of physical constants define such basic concepts as e.g. the atomic physics, the Chandrasekhar limit, black hole collapse, etc.      

The emergence of $\Lambda$ as a physical constant in modified weak field limit of GR can be tested via gravity lensing observations \cite{GS2}.

\section{Acknowledgement}
AS acknowledges the ICTP Affiliated Center program AF-04 for financial support.

\end{document}